\newcommand{\hii}{\ion{H}{2}}
\newcommand{\hei}{\ion{He}{1}}
\newcommand{\heii}{\ion{He}{2}}
\newcommand{\cii}{\ion{C}{2}}
\newcommand{\nii}{\ion{N}{2}}
\newcommand{\oii}{\ion{O}{2}}
\newcommand{\sii}{\ion{S}{2}}
\newcommand{\ciii}{\ion{C}{3}}
\newcommand{\oiii}{\ion{O}{3}}
\newcommand{\neiii}{\ion{Ne}{3}}
\newcommand{\siii}{\ion{S}{3}}
\newcommand{\feiii}{\ion{Fe}{3}}
\newcommand{\ariv}{\ion{Ar}{4}}

\newcommand{\beq}{\begin{equation}}
\newcommand{\eeq}{\end{equation}}
\documentclass[12pt,preprint]{aastex}
\shortauthors{GARNETT AND DINERSTEIN}
\shorttitle{\oii\ RECOMBINATION LINES IN THE RING NEBULA}
\received{ }
\revised{____________}
\accepted{ }
\slugcomment{To appear in the {\it Astrophysical Journal}}
\begin{document}
\title{Spatially-Resolved \oii\ Recombination Line Observations of
the Ring Nebula, NGC 6720} 
%\altaffilmark{1}}
\author{Donald R. Garnett} 
\affil{Steward Observatory, University of Arizona, 933 N. Cherry Ave., 
Tucson, AZ  85721 \\ E-mail: dgarnett@as.arizona.edu}
\author{Harriet L. Dinerstein} 
\affil{Astronomy Department, University of Texas at Austin,  
Austin, TX  78712 \\ E-mail: harriet@astro.as.utexas.edu}
\begin{abstract}
We present new long-slit CCD spectra of \oii\ permitted lines and
[\oiii ] forbidden lines in the Ring Nebula NGC 6720. These 
observations provide spatially-resolved information on both \oii\ 
and [\oiii ] over the 70$\arcsec$ diameter of the main 
shell. We find significant differences in the spatial distribution
of the \oii\ lines and [\oiii ] $\lambda$4959. The [\oiii ] emission 
follows the H$\beta$ emission measure, peaking slightly radially
inward from the H$\beta$ peak. The \oii\ emission peaks inside the
[\oiii ] emission. This suggests that radiative recombination may 
not be the primary mechanism for producing the \oii\ lines. O$^{+2}$ 
abundances derived from \oii\ lines are 5-10 times larger, than 
those derived from [\oiii ] in the region within 20$\arcsec$ of 
the central star. Outside 
of this region, however, the \oii\ and [\oiii ] derived abundances
agree to within 0.2-0.3 dex. The electron temperature derived from
[\oiii ] lines rises smoothly from about 10,000 K in the outer shell
to about 12,000 K in the center; we see no evidence for a temperature
jump that would be associated with a shock. If temperature fluctuations
are responsible for the discrepancy in O$^{+2}$ abundances, the 
average temperature would have to be approximately 6,500 K in the
He$^{+2}$ zone and about 9,000 K in the outer shell in order to
force the [\oiii ]-derived abundance to equal that derived from \oii.
This would conflict with ionization models for PNe which predict
that the temperature is higher in the He$^{+2}$ region close to
the ionizing star. We therefore argue that temperature fluctuations
can not explain the abundance discrepancy. A comparison of the spatial
distribution of \oii\ emission with the location of dusty knots
shows that the \oii\ recombination lines do not peak where the dense
knots are located, creating difficulties for models which explain the 
recombination line/forbidden line discrepancy by density fluctuations. 
We examine the possibility high-temperature dielectronic recombination 
in a central hot bubble enhances the recombination line strengths in 
the central part of the nebula. However, comparison of recombination 
rates with collisional excitation rates shows that the increase in 
recombination emission due to dielectronic recombination at $T$ 
$\approx$ 10$^5$ K is not sufficient to overcome the increase in 
collisonally-excited emission. We are unable to find a completely 
satisfactory model to explain the discrepancy between recombination 
line and forbidden line abundances.
\end{abstract}


\section{The \oii\ Recombination Line Problem in Ionized Nebulae}

For approximately the past three decades, analysis of collisionally-excited
forbidden lines in the optical/UV wavelength region has been the primary
means of determining the physical conditions and element abundances in 
\hii\ regions and planetary nebulae (PNe). Studies of these emission
lines have been a fundamental source
of data for the study of the chemical evolution of galaxies other than
the Milky Way and for comparison with models of stellar nucleosynthesis.
For almost as long a time, however, there has been the nagging question
of the reliablity of such abundance measurements. The crux of the problem
is that collisionally-excited lines in the optical/UV part of the spectrum
have large excitation energies (1-10 eV), and thus the excitation rates
are acutely sensitive to electron temperature. \citet{peimbert67} pointed 
out that if significant fluctuations about an average temperature are
present, the collisionally-excited line emission will be weighted toward
hotter regions because of the exponential dependence on $T_e$, while
the Balmer recombination lines will be preferentially produced in cooler
regions because of the inverse dependence on $T_e$. In this case, ion
abundances derived from the forbidden lines could be systematically 
underestimated by a large amount. It has not yet been demonstrated 
definitively that significant deviations from a smoothly-varying 
temperature structure exist in photoionized nebulae, other than the 
gradients expected from ionization/thermal equilibrium models. Determining 
whether such temperature fluctuations exist and their magnitude is challenging
observationally, and it is strongly debated whether temperature fluctuations
have a significant effect on abundance determinations based on 
collisionally-excited lines (see Mathis 1995 for a balanced discussion).

It is often suggested that observations of recombination lines from the
heavy elements would provide more reliable measures of element abundances
than the commonly observed collisionally-excited lines. This is 
because recombination lines have very similar dependences on electron
temperature, so the ratio of a given recombination line to, say, H$\beta$
is nearly independent of $T_e$. Thus, heavy element abundances could be
calculated from recombination lines with virtually no systematic 
uncertainties, in principle. The main drawback is that recombination lines
from elements heavier than He are intrinsically faint, typically less than
1\% of H$\beta$, because of the direct scaling of the emissivities with
ion abundance. As a result, few measurements of heavy element 
recombination lines have been published. 

Nevertheless, improved spectrographs and detectors have enabled high quality
measurements of heavy element recombination lines, particularly \oii, to be
made in recent years \citep{liu95, epte98, liu00}. At the same time,
improved computations of effective recombination coefficients for transitions
of heavy element recombination lines (e.g., Storey 1994) have made 
the analysis more reliable. For the Orion Nebula, \citet{epte98} found that
\oii\ recombination lines gave somewhat higher O$^{+2}$/H$^+$ than the
[\oiii ] collisionally-excited lines, and interpreted this in terms of
modest temperature fluctuations. On the other hand, \citet{liu95,liu00}
have found that \oii\ and [\oiii ] yield spectacularly discrepant O$^{+2}$
abundances in the planetary nebulae (PNe) NGC 7009 and NGC 6153 -- with
the \oii\ lines giving abundances 5-6 times larger than those derived from
[\oiii ]. These differences are so large that they can not be accounted
for in the context of the \citet{peimbert67} small temperature
fluctuation formulation. 

We have been obtaining new measurements of \oii\ recombination lines in
PNe to further understand the cause of the \oii\ - [\oiii ] discrepancy.
In \citet{dlg01} (hereafter Paper II), we will present measurements of
integrated \oii\ lines and derived abundances for a sample of ten PNe,
and compared these abundances with published results based on [\oiii ].
The main results from Paper II are summarized in \cite{dlg00} and 
\cite{texmex}. In particular, we find: (1) a wide range of values for 
the difference between \oii - and [\oiii ]- derived abundances 
($\Delta$log O$^{+2}$/H$^+$  
anywhere between 0.0 and 0.8 dex); (2) an inverse correlation between
$\Delta$log O$^{+2}$/H$^+$ and nebular diameter; and (3) that 
\oii\ lines arising from levels populated mainly by dielectronic 
recombination are especially enhanced.
%We also identified some intriguing correlations between 
%$\Delta$ log(O$^{+2}$/H$^+$) 
%and nebular properties that could be
%significant clues as to the origin of the discrepancies. 
In a companion program, we have obtained deep long-slit observations
of some PNe to study and compare the spatial distribution of both \oii\
and [\oiii ] emission, in the hope that this would provide new clues 
toward understanding the source of the abundance discrepancies. In 
this paper we describe the results for the largest object in our sample, 
the Ring Nebula NGC 6720.

\section{Long-slit Spectroscopy of \oii\ Lines in the Ring Nebula}

We have observed several Galactic planetary nebulae in long-slit mode
in order to study the spatial distribution of emission
from both the \oii\ recombination lines and [\oiii ] forbidden lines.
These observations
were obtained on the clear nights 6-7 June 1999 with the 2.3m Bok reflector 
at Steward Observatory, using the Boller \& Chivens spectrograph. The 
spectrograph has a 1200$\times$800 Loral CCD with 15$\mu$m pixels, 
corresponding to 0\farcs83 per pixel at the detector. The four-arcminute
long slit thus covers about 285 rows on the detector. We used an 832 
line/mm grating in second order to cover the spectral range 4150-5000 \AA.
With a 2\farcs5 slit, the spectral resolution was in the range 2.1-2.3 
\AA, with the best resolution set at approximately 4650 \AA. In our
Ring Nebula observations, the spatial profiles of stars along the slit
had a FWHM of approximately 2\farcs5. We centered the long slit on the 
NGC 6720 central star, with the slit at a position angle of 91$^\circ$, 
where the PA is measured from N through E. We made four 15-minute 
exposures at this position.

The two-dimensional data were processed by standard methods. The wavelength
solution for the spectra was determined from the positions of He-Ar lines
from calibration lamp observations taken through the same spectrograph setup. 
The root-mean-square residuals from a fourth-order polynomial solution were 
approximately 0.1 \AA. The object frames were divided by a normalized 
flat-field constructed by combining observations of the illuminated dome 
and the twilight sky. The spectra were corrected for atmospheric extinction 
using standard Kitt Peak National Observatory extinction coefficients; and 
finally, the spectra were placed on a relative photometric scale using a 
spectrometer sensitivity function derived from observations of standard stars 
from the list of \citet{mass88}. Note that because of the narrow slit, 
absolute spectrophotometry was not possible; however, the relative end-to-end 
photometric calibration is good to about 3\%; the flat-field uncertainty adds 
another 1\% to the errors.
 
After calibration, the four individual frames were combined to form a 
single image, and the sky background was subtracted. At this time we
also extracted a section of the frame 131 pixels (109\arcsec) wide 
containing only the bright emission from the nebula (along the east-west
direction), and rectified the
image to align the centroid of the central star with the rows of the
image, to ensure that features in each emission line lined up spatially. 
We saw no variation
in the background between the four separate observations, so we averaged
the frames before subtracting the background to improve the signal/noise
for the sky determination. The sky level along each column was determined
by a linear fit to the background in regions along each column outside the
bright part of the nebula. Note that the Ring Nebula has an extended halo
outside the bright inner nebula \citep{cja87}. 
The faint emission from this
halo is seen in the brightest lines in our spectrum, H$\beta$, H$\gamma$,
and [\oiii ] 4959 \AA. The halo emission extends across the entire slit;
however, this emission has a surface brightness of less than 1\% of the
surface brightness in the bright shell. Therefore, subtraction of the 
halo component will have little effect on the line ratios we derive in 
the inner nebula.

The final processing of the data involved extracting spatial cuts of
various emission lines from the image. 
%Table 1 identifies the emission
%lines extracted for our analysis. 
The extraction method was similar to
that used for extracting one-dimensional stellar spectra, except that
we did the extraction along lines of constant wavelength. As part of
the extraction we subtracted the nebular and stellar continuum by fitting
to line-free regions of continuum near the wavelengths of each extracted
emission line. The result was an array of pure emission line flux vs. 
spatial position for each extracted line. Note that continuum subtraction 
by this method does not take into account absorption lines in the central
star spectrum, or in starlight scattered by dust within the nebula, that 
may coincide with an emission line. The effect of stellar absorption will 
manifest itself as peculiar emission line ratios at the position of the 
central star, and in general we will exclude the regions including the
central star from the discussion; the effects of scatterd starlight will
be much smaller, and in any case would mean that the permitted lines are
systematically even stronger than observed. Figure 1 shows 
examples of extracted spatial profiles for a few emission lines. 
Figure 2 shows an expanded view of the spectral region around 4661 \AA\
for two different positions along the slit. Figure 2 shows that, although
[\feiii ] $\lambda$4658 is significantly stronger than \oii\ $\lambda$4661
in the inner part of the Ring Nebula, the two lines are readily 
distinguished and can be deconvolved accurately.

\section{Analysis of Emission Line Profiles}

\subsection{Interstellar Reddening}

We can estimate the interstellar reddening along the slit from the
H$\gamma$/H$\beta$ intensity ratio. This ratio depends slightly on
the nebular electron temperature $T_e$. For NGC 6720 $T_e$ is between
10,000 K and 12,000 K \citep{gmc97}; in this range of $T_e$, the 
H$\gamma$/H$\beta$ ratio
is 0.47 \citep{hs95}. The relation between
observed and reddened line ratios is given by 
\begin{equation}
{I_{obs}(\lambda)\over I_{obs}(H\beta)} =  
{I_{int}(\lambda)\over I_{int}(H\beta)} 10^{-c(f(\lambda) - f(H\beta))}, 
\end{equation}
where I$_{obs}$ and I$_{int}$ are the observed and intrinsic line intensities,
and f($\lambda$) $-$ f(H$\beta$) = 0.13 for $\lambda$ = H$\gamma$, from
the average interstellar reddening curve of \citet{sm79}. The coefficient 
$c$ (or, more commonly, $c$(H$\beta$))
is then the logarithmic `extinction' at H$\beta$ -- more accurately, a
combination of absorption and scattering.

Figure 3(a) shows the spatial variation of $c$(H$\beta$) we derive along
the slit for NGC 6720. Generally, we see values for $c$ ranging between
0.1 and 0.2 along the slit, with possibly a slow increase from east
to west along the slit. The reddening values we derive are consistent
with the values obtained by \citet{bark87} and \citet{gmc97}.
Note the spike at the position of the central star, where the line ratio 
is very likely affected by stellar absorption. Note also the steep 
upturn at the E end and the downturn at the W end. We suspect these
steep changes are not real, but artifacts resulting from small 
misalignment of the H$\beta$ and H$\gamma$ extractions, combined with
a small variation of the focus, at positions where the nebular surface
brightness profile is declining rapidly (see Figure 1). 

Nevertheless, the correction for reddening relative to H$\beta$ is small 
for all lines in our observed spectral region along most of the slit, less 
than 10\% over the region $\pm$38\arcsec\ from the central star. 
This is illustrated in Figure 3(b), where we show the ratio of
I(H$\gamma$)/0.47~I(H$\beta$). If there were no reddening, this ratio
would be close to unity; Figure 3(b) shows that this ratio is within 10\% 
of unity over the region where the emission is bright. If we ratio lines 
in the blue half of the spectrum to H$\gamma$ and those in the redward 
half to H$\beta$, the error accrued is less than 5\% over the regions of 
interest in the nebula. In the end, we chose this method to correct the 
line intensities. The regions of the nebula beyond $\pm$40\arcsec\ from 
the central stars have too low S/N for the faint lines to contribute to 
our analysis, so we will exclude them from the following discussion. 

\subsection{Electron Density and Electron Temperature}

Our spectrum includes the [\ariv ] 4711, 4713 \AA\ doublet; the ratio
of these two lines is sensitive to the electron density $n_e$, similar 
to the [\sii ] 6717, 6731 \AA\ pair. There is one complication in using
the [\ariv ] ratio: the 4711 \AA\ line is blended with the \hei\ 4713 \AA\
line at low resolution. This is apparent in our spectrum by a larger
line width for the 4711 \AA\ line, as well as a shift in wavelength
from the inner nebula, where [\ariv ] is strong and \hei\ weak, to the
outer ring where \hei\ predominates. 

We can correct for the \hei\ 4713 \AA\ contamination by subtracting a
scaled version of the \hei\ 4471 \AA\ line profile. \citet{gmc97} 
estimated $n_e$ from [\sii ] line ratios in the Ring, obtaining
values of 200-600 cm$^{-3}$. At these densities, collisional excitation
has a small effect on the 4713/4471 line ratio, so we choose to neglect 
it. For $T_e$ $\approx$ 10,000 K, the 4713/4471 intensity 
ratio is 0.11, based on the emissivities computed by \citet{bss99}. 
Assuming a constant temperature in the
\hei\ zone, we multiply the 4471 \AA\ fluxes by 0.11, then subtract
the result from the 4711 \AA\ line fluxes. The correction to the 
4711 \AA\ fluxes are small for the inner nebula, where \hei\ emission
is faint, and greatest in the outermost parts of the nebula. 

We estimated $n_e$ from two regions on either side of the central star, 
using the atomic code described by \citet{shawduf95}, assuming an electron 
temperature $T_e$ = 11,000 K. East of the central star, the mean [\ariv ] 
ratio was 1.33$\pm$0.03 (where the error is the rms scatter about the mean), 
which corresponds to $n_e$ = 740$\pm$250 cm$^{-3}$, while west of the
central star, the [\ariv ] ratio was 1.34$\pm$0.06, which yields $n_e$ = 
660$\pm$500 cm$^{-3}$. These values are consistent with the densities 
obtained by \cite{lp94}, \cite{lb96} and \citet{gmc97}. \citet{bark87} 
derived densities approximately twice as large as ours; however, this is 
the result of using atomic data for [\sii ] that are now out of date. We 
have re-computed $n_e$ using Barker's data and the Shaw \& Dufour program, 
and obtain densities similar to the ones we obtain from [\ariv ]. 

Since we observe both [\oiii ] 4363 \AA\ and [\oiii ] 4959 \AA\, we can
estimate the electron temperature $T_e$ across the nebula. 
%We show the variation of the I(4363)/I(4959) ratio in Figure 5. 
%The ratio exhibits a slow increase from the outer shell to the interior. 
Figure 4(a) plots
the variation of $T_e$ derived from the [\oiii ] ratio, showing a
smooth increase in $T_e$ from the shell to the interior of the nebula,
from about 10,000 K in the shell to about 12,000 K in the center. This
behavior is in good agreement with that derived by \citet{gmc97}.

For comparison, we plot in Figure 4(b) the [\oiii ] $\lambda$4959/H$\beta$ 
line ratio across the slit. We note several features in the plot. First,
the line ratio shows only a slow variation across most of the nebula,
with a shallow minimum in the center. Second, we see a steep decline
in [\oiii ]/H$\beta$ at about 36\arcsec\ from the central star in 
both directions. This most likely corresponds 
to the transition from O$^{+2}$ to O$^+$ near the edge of the bright
shell. Third, [\oiii ]/H$\beta$ increases again outside the bright shell,
corresponding to the inner halo region, which evidently has rather high
excitation and/or high ionization along these sightlines. 

Note that the minima in $T_e$ do not correspond to the minima in 
[\oiii ]/H$\beta$, but rather lie just interior to the low-excitation
zones. $T_e$ then appears to increase a bit in the region of low 
[\oiii ]/H$\beta$; this increase may be a result of radiation hardening
combined with inefficient cooling in the low-ionization zone of the
nebula. The signal/noise in our 4363 \AA\ line observation is too low
to determine $T_e$ in the halo region. 

\subsection{Determination of Ion Abundances Across the Nebula}

Now that we have established the variation of $T_e$ across NGC 6720,
we can proceed to determine abundances for the ions of interest. For
this study, we will derive abundances relative to H$^+$ for O$^{+2}$
(from both [\oiii ] 4959 \AA\ and \oii\ 4661 \AA), 
C$^{+2}$ (from \cii\ 4267 \AA), 
He$^+$, and He$^{+2}$, as a function of position along our slit.
Since the electron densities across the nebula are relatively small,
we can assume the low-density limit and compute the ion abundances
analytically.

The abundance of any ion relative to H$^+$ derived from the ratio of
the intensity of a transition $\lambda$ to the intensity of H$\beta$ 
is given by
\begin{equation}
{N(X^{+i})\over N(H^+)} = {I(\lambda)\over I(H\beta)}{\epsilon(H\beta)\over
\epsilon(\lambda)},
\end{equation}
where $\epsilon(\lambda)$ represents the volume emission coefficient
for a given emission line $\lambda$. For collisionally-excited lines in 
the low-density limit, the analysis in section 5.9 of \citet{osterb89} 
applies. 

The volume emission coefficient for a collisionally-excited line 
is given by
\begin{equation}
\epsilon(\lambda) = h\nu q_{coll}(\lambda) = {hc\over \lambda}~8.63\times10^{-6}(\Omega/\omega_1)T_e^{-0.5}e^{-\chi/kT_e} 
\end{equation}
where $\Omega$ is the collision strength for the transitions observed,
$\omega_1$ is the statistical weight of the lower level, and $\chi$ is 
the excitation energy of the upper level. For [\oiii ] we use the 
collision strengths from \citet{lb94}, interpolated to a temperature 
of 11,000 K; the collision strength for the 4959, 5007 \AA\ lines 
varies by only 8\% between 10,000 K and 15,000 K, so the assumption 
of a constant $\Omega$ introduces only a small error.

For recombination lines, the emission coefficient is given by
\begin{equation}
\epsilon(\lambda) = h\nu q_{rec}(\lambda) = {hc\over \lambda}~\alpha_{eff}(\lambda),
\end{equation}
where $\alpha_{eff}$($\lambda$) is the effective recombination coefficient
for the recombination line $\lambda$. For the \oii\ 4661 \AA\ line, 
we use the effective recombination coefficient derived by 
\citet{storey94} for LS-coupling in Case B. For H and \heii\ we 
used the recombination coefficients from 
\citet{hs95}, while for \hei\ we used the emissivities computed
by \citet{bss99}. We employ a value 
$\alpha_{eff}$(\oii\ $\lambda$4661)/$\alpha_{eff}$(H$\beta$) = 1.29 for 
$T_e$ = 10,000 K; this ratio varies by only 4\% between 10,000 K and 
15,000 K. Similarly, $\epsilon$(\heii\ $\lambda$4686) = 11.9$\epsilon$(H$\beta$)
for $T_e$ = 11,000 K, varying by only 9\% between 10,000 K and 20,000 K,
and $\epsilon$(\hei\ $\lambda$4471) = 0.501$\epsilon$(H$\beta$) at $T_e$ =
10,000 K and $n_e$ = 100 cm$^{-3}$; collisional excitation increases
$\epsilon$($\lambda$4471) by only 5\% at $n_e$ = 10$^4$ cm$^{-3}$
\citep{bss99}. Given the relatively low 
densities we derive for the Ring, and the small variation across the
nebula, we can neglect collisional excitation of \hei.
Finally, we computed C$^{+2}$ abundances from the \cii\ 4267 \AA\ line
using the Case B effective recombination coefficient from \citet{davey00} 
and \citet{liu00}. At $T_e$ = 10,000 K, 
$\alpha_{eff}$(\cii~$\lambda$4267)/$\alpha_{eff}$(H$\beta$) $\approx$ 9.1. 

\section{Results}

Figure 5 shows a comparison of the spatial distribution of the [\oiii ]
4959 \AA\ emission and the \oii\ 4661 \AA\ emission in the Ring. It is
readily apparent that the [\oiii ] and \oii\ lines have distinctly 
different spatial distributions. The [\oiii ] emission essentially follows
the H$\beta$ emission, peaking slightly interior to the H$\beta$ peak,
consistent with the ring-like morphology of the nebula. In contrast, 
the \oii\ emission peaks $interior~to$ the [\oiii ]-emitting region. This 
behavior is remarkable - if the \oii\ $\lambda$4661 emission arises mainly 
from radiative recombination of O$^{+2}$ to O$^+$, then one would 
predict that the \oii\ emission would peak $farther$ from the central 
star than the [\oiii ] emission, where the ionization decreases. Clearly, 
the \oii\ and [\oiii ] emission lines arise under dissimilar conditions. 

In Figure 6 we show the computed O$^{+2}$ abundances as derived from
the forbidden lines and the recombination lines, using the electron
temperatures shown in Figure 4(a). Again, we see that 
O$^{+2}$ as derived from [\oiii ] peaks in the shell while O$^{+2}$
derived from the recombination line is more concentrated toward the
center of the nebula. To demonstrate that the results are not a result
of low signal/noise in the 4661 \AA\ line, we also plot in Figure 6
the abundances we obtain from binned (summed over 8-12\arcsec) spectra 
extracted at several intervals along the slit. The agreement between
the binned spectra and the point-by-point measurements is excellent. 

Another thing that is apparent upon inspection of Figure 6 is that
the discrepancy between the [\oiii ] and \oii\ abundance results is
not constant; the discrepancy is much larger for the central regions 
of the nebula, where O$^{+2}$(\oii) is as much as ten times larger
than O$^{+2}$([\oiii ]), than in the bright shell, where the difference
is not more than a factor of two. Similar behavior has also been 
seen by \citet{liu00} in the planetary nebula NGC 6153, another 
shell-like nebula. This suggests a common origin for the 
forbidden-line/recombination-line abundance discrepancy.

The helium abundance derived from our observations is shown in Figure 
7, with the total He/H plotted as the solid line and the He$^{+2}$
contribution as the dashed lines. The plot shows that the helium
abundance remains rather constant across the Ring Nebula, apart from
possibly a small upturn at the edges of the shell. This increase is
seen in He$^+$, and may be due to the assumption of a constant electron
density across the shell; however, the effect is not large. The average
helium abundance we derive, He/H $\approx$ 0.12, is about 10\% smaller
than that derived by \citet{gmc97}, He/H $\approx$ 0.135. The reason
for the discrepancy is not clear. We do agree, however, that the 
helium abundance is essentially constant across the Ring.

Finally, in Figure 8 we display the C$^{+2}$ abundances derived
from our measurements of \cii\ $\lambda$4267, compared with the
O$^{+2}$ derived from \oii. The C$^{+2}$ abundance shows an increase
toward the nebula center similar to that of O$^{+2}$, although the
increase is not as strong. We find that C$^{+2}$/O$^{+2}$ $\approx$ 1
in the bright shell, decreasing to $\approx$ 0.5 (approximately the
solar value)
in the center. This variation could be due to ionization effects 
(increasing importance of C$^{+3}$ in the interior), a real C/O 
abundance variation across the nebula, or a variation in the emission
mechanism. Note that the C$^{+2}$ abundance in the shell is similar
to that derived for NGC 6720 (C/H $\approx$ 6$\times$10$^{-4}$) from
the \ciii ] 1908 \AA\ feature by \citet{kwithen98}.

\cite{bark87} derived C$^{+2}$ abundances in the Ring Nebula from
measurements of \cii\ $\lambda$4267 and from $IUE$ measurements of
\ciii ] $\lambda$1908. He found that the abundances derived from
these two lines differed by a large factor in the central part of
the Ring Nebula, but that the difference decreased radially outward
and essentially disappeared in the bright shell. This behavior is
very similar to what we see for abundances derived from \oii\ and
[\oiii ]. This suggests a common mechanism affecting emission lines
from both O$^{+2}$ and C$^{+2}$. 

\section{Discussion}

Our long-slit observations of the Ring Nebula demonstrate that the 
\oii\ and [\oiii ] lines have very different spatial distributions. 
The big surprise is that the peak recombination line emission occurs
in the region interior to the [\oiii ] zone, in contrast to the 
expectations if the \oii\ emission is the result of radiative 
recombination of O$^{+2}$. Either the \oii\ lines are enhanced in the 
inner nebula or the electron temperature derived from [\oiii ] is 
greatly overestimated in the nebular interior. 

In Paper II, we identify other surprising results. We find that the size 
of the \oii -[\oiii ] abundance discrepancy, 
$\Delta$ log(O$^{+2}$/H$^+$),
is anticorrelated with PN diameter and mean emission measure. We also see 
that \oii\ lines that arise from levels populated mostly by dielectronic 
recombination are more discrepant than other \oii\ lines when compared with 
the forbidden lines. These results, together with our new observations of 
the Ring Nebula, must be giving us important clues regarding the origin of 
the discrepancy between the recombination lines and the forbidden lines. 
Here we discuss some of the possible mechanisms for producing the abundance 
discrepancies; other possibilities will be examined in Paper II.

\subsection{Temperature Fluctuations}

Temperature fluctuations have been invoked to explain why 
collisionally-excited lines give lower abundances than the recombination 
lines \citep{pst93}. Nevertheless, the abundance discrepancies observed 
in NGC 7009 \citep{liu95}, NGC 6153 \citep{liu00}, and the Ring Nebula
(this paper) are so large that they are are no longer in the regime
where Peimbert's formalism of small temperature fluctuations applies. For
example, in the Ring Nebula we find that the \oii\ lines yield abundances
as much as ten times larger than the [\oiii ] lines in the center of the
nebula. If we assume that the \oii\ lines give the correct abundances,
we can estimate the magnitude of temperature fluctuations needed to 
obtain the same abundances from the [\oiii ] lines. Forcing the [\oiii ] 
lines to give the same abundances requires an average temperature
$T_0$ $\approx$ 6,500 K in the center of the nebula, and $T_0$ $\approx$
9,000 K in the shell, whereas we derived $T$[\oiii ] $\approx$ 12,000 K 
in the center and $T$[\oiii ] $\approx$ 10,000 K in the shell. Using
the $t^2$ formalism of \citet{peimbert67}, these temperature differences
imply $t^2$ $\approx$ 0.03 in the shell and $t^2$ $\approx$ 0.16 in the
center of the nebula! $t^2$ = 0.16 corresponds to rms temperature 
variations of $\pm$40\%. It is difficult to imagine how such wild
temperature variations could go unobserved [see Fig. 4(a)]. 

Still, we saw in Figure 4(a) that there is a radial gradient in T[\oiii ]
within the Ring Nebula. This variation could give rise to a noticeable
$t^2$ term if the temperature measured from the summed spectrum differs 
significantly 
from the emission-measure-weighted average temperature. We compare
the [\oiii ] temperature derived from the Ring Nebula spectrum summed 
along the entire slit with the average [\oiii] temperature weighted by 
the [\oiii ] $\lambda$4959 intensity along the slit. The temperature 
derived from the full-slit spectrum is 10,950 K, while the emission-weighted
average temperature is 10,860 K. Using equation (4) of \cite{gar92},
the measured difference in T[\oiii ] corresponds to $t^2$ = 0.003. This
is too small to have a significant impact on abundances. Nevertheless,
such gradients in $T_e$ are likely to be common in \hii\ regions and
PNe, and should be taken into account whenever possible.

\citet{liu00} have also argued that temperature fluctuations are unable
to account for the large discrepancy between forbidden-line and 
recombination-line abundances in NGC 6153. They derived $T_e$ from
both the [\oiii ] lines and from the Balmer jump/line ratio in NGC 6153; 
the Balmer jump temperature was determined to be 3000 K smaller than the 
[\oiii ] 
temperature. Nevertheless, \citet{liu00} found that this difference in 
$T_e$ could account for only one-third of the factor of nine difference 
in the derived O$^{+2}$ abundances in NGC 6153. Furthermore, they found 
that abundances derived from $ISO$ measurements of collisionally-excited
[\oiii ], [\neiii ], and [\siii ] IR fine-structure lines agree very well
with the results from corresponding optical forbidden lines. Since the
IR fine-structure lines have a very weak temperature dependence compared 
to the optical forbidden lines, it is difficult to reconcile the good 
agreement in IR and optical abundances with the idea of significant 
temperature fluctuations. These results and ours are evidence that 
temperature fluctuations can not account entirely for the large 
differences between recombination lines and forbidden lines in PNe.

\subsection{Shocks}

Shocks can provide a source of additional heating for a photoionized gas,
and have been proposed as a mechanism to account for temperature fluctuations 
\citep{psf91}. Shocks can heat a photoionized plasma by transforming the 
kinetic energy of a flow (e.g., a stellar wind) into random thermal energy, 
and strong shocks heat the ambient medium to very high temperatures. 
High-excitation lines such as [\oiii ] 4363 \AA\ would be preferentially 
enhanced behind shocks in photoionized nebulae. \citet{psf91} showed that 
an interstellar shock superposed on the spectrum of an \hii\ region could 
lead to a spuriously high measured electron temperature, and thus to 
underestimated abundances from collisionally-excited lines
in the optical spectrum. 

In a PN, a shock front is expected to be formed at the interface between 
a fast stellar wind from the central star and the slowly-expanding circumstellar
envelope. The fast stellar wind would carve out a hot bubble in the interior
of the nebula which would emit primarily in X-rays. Clear evidence for such 
hot bubbles has been found recently for a small number of PNe from X-ray 
imaging \citep{ksvd00, kvs01, chu01}. 

Evidence that shocks enhance temperatures in photoionized gas in an
observable way is generally lacking, 
however. Physically, a shock is an extremely thin structure, only 
a few particle mean free paths thick; the high-temperature recombination
zone downstream from the shock is generally only a few times 10$^{15}$ cm
deep in interstellar shock models \citep{shmck79, raymond79, bdt85}. 
Therefore, one might expect the shock front to be manifested by a sharp, 
narrow spike in the spatial profile of $T_e$. Examination of Figure 4
shows that we observe no sharp jumps in $T_e$ across the Ring, only a
smooth increase from the shell to the center. The rms deviations from
this smooth profile are less than 200 K. One must consider that the 
spatial resolution of our data, 2$\farcs$5 $\approx$ 2$\times$10$^{16}$
cm at a distance of 600 pc, is larger than the expected width of the
shock cooling zone, and that we are observing a temperature averaged 
along a chord through the nebula. 

Another problem that must be considered is the structure of the PN and
its effect on the propagation of the shock. \citet{frank94} has simulated
the radiation-gas dynamical evolution of a PN in the context of the 
interacting-winds model \citep{kpf78, kwok94}. These models begin 
with a slowly-expanding
red giant wind. The PN expansion is initiated by the onset of a slow,
high-density ``superwind'' which lasts a few thousand years, creating
a dense shell of gas around the central star. Following the end of the
superwind, a high-velocity, low-density ``fast wind'' ensues, clearing
out a central cavity and exposing the hot stellar core. Frank followed
the gas dynamics and ionization/thermal evolution of the nebula 
simultaneously.

In the model, the fast wind drives a shock into the superwind, creating
a hot central bubble. The low density in the bubble allows radiation from
the hot stellar remnant to penetrate and ionize the dense shell. The high 
pressure in the hot bubble drives a forward shock into the dense shell. 
Figure 2 of \citet{frank94} shows a snapshot of the dynamical, thermal,
and ionization properties of the nebula at a point which is comparable to 
the current state of the Ring Nebula. At this stage, the nebula is fully 
ionized, and the temperature profile in the dense shell is determined by 
radiative cooling and heating. At this point, the forward shock is nearly
isothermal and does not produce a noticeable temperature jump. This 
behavior is consistent with the lack of a temperature discontinuity in
our observations of the Ring Nebula. However, a sharp jump in $T$ is
seen at the contact discontinuity between the hot bubble and the dense
shell in Frank's models. One might then expect to see sharp jump in 
observed $T_e$ at this point in a PN. We do not observe such a jump in
the Ring, but this may be due to insufficient spatial resolution. 

Considering the results of our observations and Frank's models leads us
to conclude that shocks can not explain the large discrepancy between
forbidden-line and recombination-line abundances, although shocks should
be present. Spectroscopic observations with better spatial resolution
should show whether shocks produce any significant effect on temperature
profiles in PNe.

\subsection{Abundance Inhomogeneities}

\citet{tpp90} and \citet{peimbert93} suggested that abundance inhomogeneities
could produce anomalously strong recombination lines from heavy elements in 
PNe. If the central region of the PN contains hydrogen-depleted material, 
the metal-rich interior would be expected to have a low electron temperature 
due to strong cooling by IR fine-structure line emission. Thus, optical 
forbidden lines would be weakened in this region while the recombination 
lines are enhanced. The optical forbidden lines would be produced mainly 
in the H-rich shell and would reflect the conditions and abundances there. 

A number of observable consequences result from this model. First, it
might be expected that the central nebula would be enriched in helium
as well, if the C and O enhancements result from incomplete mixing of
He-burning products. Figure 7, however, shows that this is not the case for 
the Ring Nebula -- He/H is essentially constant across the nebula. 
Alternatively, one might hypothesize that the central region is enhanced
in C and O by a wind from a H- and He- depleted central star. However, the 
central star of the Ring Nebula is classified as H-rich \citep{nap95, nap99}. 

Second, one would expect the observed $T_e$ to decrease toward the center
of the nebula. From Figure 4, however, we see that $T_e$ $increases$
toward the center. One could get around this by arguing that the derived
temperature toward the center does not reflect the conditions in the
interior, but rather in the shell projected along the line of sight.
\footnotemark
\footnotetext{Another possible argument is that the higher temperature
in the central nebula is the result of shocks located preferentially 
along those lines of sight. The kinematics discussed in \citet{gmc97} 
appear to rule out such a situation.}
Such a condition could occur if the nebula were shaped like a lozenge
or an ellipsoid with the minor axis along the line of sight to the 
central star. In this configuration, gas near the minor axis would 
be closer to the central star and be subject to a stronger ionizing 
radiation field (i.e., a higher ionization parameter), and thus would
have a higher $T_e$. This configuration conflicts with the model for
the Ring presented by \citet{gmc97}. Based on the measured velocity
field in the nebula, they argue that the Ring is a prolate ellipsoid
with the major axis pointing at a small angle to the line of sight, a 
configuration roughly orthogonal to the one discussed above.

Finally, IR fine-structure line measurements do not show much evidence 
for the large abundance enhancements needed to explain the observed 
recombination line emission \citep{dhew95,liu00}. (See also the 
discussion in section 5.1.) From the above discussion, we conclude 
that the strong recombination line emission in the central region of 
the Ring does not reflect an H-poor, metal-enriched interior.

\citet{liu00} have also considered the question of abundance inhomogeneities.
They first determined that simple density inhomogeneities in a nebula with 
homogenous
abundances could not account consistently for the strengths of forbidden 
lines, recombination lines, and the Balmer line ratios in NGC 6153.
They next considered super-metal-rich (SMR) condensations, and found that 
they could account for the emission line strengths in NGC 6153 with a 
model having dense, H-depleted inclusions that occupy a very small fraction 
of the volume. In this model, the heavy element recombination lines arise 
predominantly in the cool, dense condensations, while the forbidden lines 
are emitted primarily in the hotter, low-density component. \citet{liu00}
were able to produce models of this kind that were able to reproduce the 
strengths of the recombination lines as well as the optical, UV, and IR
collisionally-excited lines. 

Chemical inhomogeneities might be expected in a PN, since C, N, and He
are produced in different regions of the progenitor star. Convective 
mixing during the red giant phase could homogenize the PN envelope, but
such mixing might never be complete, especially in very late stages 
prior to envelope ejection. Hydrodynamic instabilities would promote
clumping in the gas, and a clumpy stellar wind may eject chemically
inhomogeneous knots of gas. Evidence for knots of H-depleted material is 
seen in several PNe, for example Abell 30 \citep{jac79}.

On the other hand, there are a number of difficulties with this model
as an explanation for the apparent abundance discrepancies. \citet{liu00}
noted the difficulty of maintaining the postulated dense knots in 
pressure equilibrium with the surrounding nebula. Another difficulty 
is the roughly equal enhancements of C, N, O, and Ne in the recombination 
line emission. This is inconsistent with the nucleosynthesis predicted 
for PN progenitors, which generally synthesize C, N, and He, but not O 
or Ne. 

In Paper II, we show that the difference between forbidden-line and 
recombination-line abundances in PNe is inversely correlated with the 
nebular emission measure and diameter: in compact, high-surface-brightness 
PNe the forbidden lines and recombination lines give essentially identical
abundances, while large, low-surface-brightness PNe show large abundance
discrepancies. This result suggests that the discrepancy between forbidden 
lines and recombination lines is apparent only when a PN is large and
evolved. It is difficult to imagine why hydrogen-depleted knots would 
condense out of the PN only at late times. If hydrogen-depleted clumps 
were present in the ejected envelope one would expect 
them to be present at early times as well. Evaporation of grains from 
dense globules can not produce the SMR clumps required -- enhanced 
recombination lines are seen not only for O and C, but also for N and Ne, 
which largely do not condense onto grains. One possibility is a clumpy 
stellar wind from an H-poor 
central star, and indeed, the central star of NGC 6153 shows O VI emission 
indicating an H-depleted wind. However, as noted above, the Ring 
Nebula central star is H-rich. Obviously, a better understanding of the 
ejection and clumping of material in the nebula is needed to 
determine if this model can successfully account for PN emission.

An obvious consequence of the SMR knot model is that one might expect to 
see strong \oii\ recombination lines associated with the presence of dense 
knots.  Numerous small cometary knots have been observed in the Helix 
Nebula NGC 7293 \citep{vv68}, and have been identified as elongated,
dense, dusty globules with a surface photoilluminated by the central
star \citep{odh96}. Their origin has not been determined yet, and
spectroscopic data on physical conditions and metallicities are not
yet available. However, they (or the shadow regions behind them) are
logical places to look for emission anomalies, such as enhanced 
recombination line emission. 

Imaging with the {\it Hubble Space Telescope} has revealed the presence
of dusty, possibly cometary, knots in the Ring Nebula as well. We have 
extracted WFPC-2 images of the Ring Nebula from the {\it HST} archive 
(taken under program GO-7632). We obtained images in F502N 
([\oiii ]$\lambda$5007), F658N ([\nii ] $\lambda$6584), and F469N 
(\heii\ $\lambda$4686). The three individual exposures in each filter 
were combined and mosaiced after removing cosmic rays and image defects. 
We combined the final filter images to highlight the dusty globules. 
(The STScI Office of Public Outreach web site oposite.stsci.edu has a 
nice color image of the Ring Nebula which clearly shows the dusty knots.) 
We rotated the image so that the E-W direction was aligned with the 
horizontal axis, and overlaid our spectrograph slit on the image. The 
upper panel of Figure 9 shows a negative composite image of NGC 6720 in 
the F469N, F502N, and F658N filters, with our slit outlined. Numerous 
dusty knots are seen in this image, beginning about 20$\arcsec$ from 
the central star. High-resolution IR imaging of NGC 6720 shows that 
bright H$_2$ S(1) emission follows the rings of dusty knots \citep{kas94}, 
indicating the presence of neutral and molecular material as well.
\notetoeditor{PLEASE MAKE FIGURE 9 A FULL-PAGE FIGURE.}

The lower panel in Figure 9 shows the distribution of O$^{+2}$/H$^+$
across this section of the Ring. Comparison with the upper panel shows
that the peak of the recombination line emission does not coincide
with the region of dusty knots, but rather lies interior to the rings of 
knots. The discrepancy between the abundance derived from \oii\ and 
that derived from [\oiii ] is largest in the innermost regions and 
is declining in the regions where the dusty knots are seen. Although 
the spatial resolution of our spectrum is poor compared to the {\it 
HST} imaging, we conclude that the knots observed in the Ring are not 
the primary source for the enhanced recombination emission. Any 
metal-rich knots of the kind proposed by \citet{liu00} must be 
largely invisible at {\it HST} resolution. 

\subsection{Enhanced Dielectronic Recombination}

We show in Paper II (also Dinerstein, Lafon \& Garnett 2000 and Garnett 
\& Dinerstein 2001) that some \oii\ lines yield much larger O$^{+2}$
abundances than other lines. In particular, the 4590, 4596 \AA\ lines
from multiplet 15 are especially strong. The levels that these lines
arise from are populated mainly by dielectronic recombination (DR). 
The large abundances obtained from these lines, even after taking 
into account the low-temperature dielectronic recombination rates 
\citep{ns84}, indicate that low-temperature dielectronic recombination 
at nebular temperatures does not account 
for the great strength of the lines. We therefore investigate the 
possibility that the dielectronic recombination rates in the affected
PNe could be severely underestimated. This could be the case if the
rate coefficients to upper levels are grossly underestimated (which
does not seem very likely for all of the ions involved), or if there
is a component of dielectronic recombination that is not accounted
for in standard PN emission models. We hypothesize that this could
be high-temperature dielectronic recombination in a hot central bubble;
dielectronic recombination rates are enhanced by as much as a factor
of ten at $T$ $\approx$ 10$^5$ K compared to the rates at nebular
temperatures of $\approx$ 10$^4$ K \citep{np97, nahar99}. 

The spatial distribution of \oii\ emission in our long-slit observations 
the Ring Nebula would appear to be consistent with this hypothesis. The 
\oii\ lines are strongest just interior to the main shell, in the 
region where the interface between a hot bubble and the denser ionized 
shell would be expected. The spatial profile of the \oii\ emission 
peak is fairly broad, perhaps broader than might be expected from 
such a transition layer; however, we have not accounted for the relatively 
low spatial resolution of the spectra nor for the effects of nebular 
geometry. Unfortunately, no X-ray images of the Ring Nebula exist, 
so we have no information on the presence or absence of hot gas. 

At what temperature would the recombination coefficient be large enough
to account for the increased emission? Let us assume that the \oii\ emission
in the main shell represents the intrinsic O$^{+2}$ abundance of 
(7$\pm$1)$\times$10$^{-4}$, at a temperature of 10,000 K. The maximum
derived O$^{+2}$ abundance is (24$\pm$2)$\times$10$^{-4}$, or about 
3.5 times the abundance in the shell. Equality of abundance would then
imply that the recombination coefficient in the inner nebula is higher
by the same factor. Based on the theoretical recombination coefficients
tabulated by \citet{nahar99}, this is true if the material producing 
the excess \oii\ emission is at $T$ $\approx$ 65,000 K. Similarly, the 
C$^{+2}$ abundance
in the inner nebula is about twice as large as that in the shell. The
recombination rate for \ciii\ $\rightarrow$ \cii\ is a factor two larger 
at $T$ $\approx$ 35,000 K than the rate at 10,000 K. On the other hand,
\citet{savin00} points out that the high-$T$ DR rate from these LS-coupling 
calculations may be underestimated by perhaps a factor of 1.5-2 because 
of missing autoionizing levels, which would act in favor of our model. 

One difficulty that arises is that any O$^{+2}$ ions recombining in
this hot zone would be expected to emit very strongly in [\oiii ] and 
a large increase in $T_e$ and in the forbidden line strengths might be 
observed. This problem could be mitigated if the density in the hot 
zone is very small; the contribution to [\oiii ] emission from the hot 
zone would then be overwhelmed by the strong emission from the main shell. 
We can check this by comparing the collisional excitation rates for 
[\oiii ] with the recombination rates as a function of temperature. For
the enhanced DR model to work, two conditions must be satisfied: first,
there must be a temperature regime in which the recombination rate 
$q_{rec}$ increases faster with increasing $T_e$ than the collisional 
excitation rate $q_{coll}$, at temperatures where DR dominates; and 
second, there must be a temperature regime in which $q_{rec}$/$q_{coll}$ 
exceeds the value of this ratio at 10,000 K, the electron temperature 
we measure in the shell of the ring.

Figure 10 shows the results of a comparison of recombination rates to
collisonal excitation rates for O$^{+2}$, where we plot $q_{rec}/q_{coll}$
vs temperature for both [\oiii] $\lambda\lambda$4959,5007 and $\lambda$4363.
Here we have computed the total recombination rate and assumed that the
effective recombination rates to individual levels scale identically.
We used the radiative recombination rates formulated by \citet{ppb91},
together with the dielectronic recombination rates from \citet{ns83} 
and \citet{bp89} to estimate $q_{rec}$. The collisional rates 
$q_{coll}(\lambda)$ were computed using the averaged temperature-dependent 
collision strengths of \citet{bbk81}, \citet{agg83}, and \citet{lb94}. 
Figure 10 shows that while the first condition listed above is met, the 
second is not. The plot shows that $q_{rec}$ does increase faster than 
$q_{coll}$ in the range 4.4 $<$ log $T$ $<$ 5.5.  
On the other hand, while we do see that 
$q_{rec}$/$q_{coll}$($\lambda\lambda$4959,5007)
for log $T$ = 5.1-5.7 exceeds the ratio at 10$^4$ K, the largest
difference is only about 0.1 dex. This is not nearly enough to explain
the large difference between recombination line abundances and forbidden
line abundances in NGC 6720. Furthermore, Figure 10 shows us that for 
log $T$ $>$ 4.5 $q_{rec}$/$q_{coll}$($\lambda$4363) never exceeds the value at
10$^4$ K. In order for this model to work, the theoretical DR rates would 
have to be increased by a large factor, which appears to be outside the
estimated uncertainties in current DR rate calculations. We are thus 
forced at the present time to conclude that the recombination/collisonal 
excitation ratio can not be increased sufficiently by the presence of hot 
gas, for any combination of gas temperatures, to account for the observations, 
at least with conventional physics.
%The model would then require a higher 
%$T$ in the hot zone with a higher DR rate. The maximum DR rate for \oiii 
%in the \citet{nahar99} calculations, at $T$ $\approx$ 160,000 K, is about 
%six times the rate at 10$^4$ K; 
%Even at this increased DR rate, it may not be possible to reduce
%the density enough to avoid seeing the effects of the hot gas on [\oiii ].
%On the other hand, it is possible that the observed increase in T[\oiii ]
%toward the center of the Ring Nebula (and in NGC 6153 as well) could be
%due to the effects of a hot bubble/dense shell interface, smoothed out
%by poor spatial resolution and geometric effects. A model for the emission
%profile of a PN with a hot bubble interacting with the ionized shell 
%(beyond the scope of this paper) is needed to investigate the viability
%of this hypothesis. 

\section{Summary and Conclusions}

We have demonstrated in this paper that the \oii\ and [\oiii ] lines
in the Ring Nebula have distinctly different spatial distributions 
in the nebula; similar results have been found by \citet{liu00} for
NGC 6153. The permitted line emission peaks closer to the central 
star than the forbidden line emission. This spatial distribution appears
to be inconsistent with a model in which the permitted lines arise 
mainly from radiative recombination of O$^{+2}$ in the transition zone
to O$^+$ in the bright shell. In 
order to force the abundances from the [\oiii ] lines to agree with those 
derived from \oii\ would require such large temperature fluctuations that 
it would be difficult to hide them. Furthermore, it is difficult to 
reconcile large temperature fluctuations with the good agreement between 
optical and far-IR measurements of forbidden lines in NGC 6153. Thus, it 
appears clear that temperature fluctuations can not account for most of 
the discrepancy between forbidden-line and recombination-line abundances
in planetary nebulae.

Nor do we find that C$^{+2}$ and O$^{+2}$ derived from the optical
recombination lines have similar overabundances relative to those
obtained from the collisionally-excited lines in the Ring Nebula. 
Comparing our C$^{+2}$ abundance with that obtained by \cite{kwithen98}
from the UV \ciii ] lines, we find that the RL abundance for C$^{+2}$ 
is about a factor of two higher, while for O$^{+2}$ the RLs give an 
abundance that is a factor of four higher than that derived from [\oiii ]. 
Thus it appears that the cause of the abundance discrepancy need not 
apply equally to all of the species.

Based on the enhanced strengths of \oii\ multiplet 15, the correlation 
of the abundance discrepancy with nebular size (Paper II, Garnett \&
Dinerstein 2001), and the spatial distribution of the \oii\ emission, 
we proposed that dielectronic recombination in a hot gas bubble might 
greatly enhance the emission rates of the permitted lines above the 
rates at nebular temperatures. A comparison of recombination rates 
versus collisional excitation rates for O$^{+2}$ showed, however, that 
the increase in recombination rates due to dielectronic recombination 
at $T_e$ $\approx$ 10$^5$ K is not enough to overcome the increase in 
collisionally-excited emission. Unless the theoretical DR rates are 
greatly underestimated, this does not appear to be a viable explanation 
for the recombination line - forbidden line discrepancy. In any case, 
our results suggest that we need to understand better the effects of 
DR on recombination line strengths in general.

A model in which the recombination emission arises mainly in cold, 
H-depleted clumps \citep{liu00} could explain the enhanced recombination 
line emission in PNe, although we are unable to find a close correlation 
between enhanced \oii\ emission and the locations of dusty globules in 
the Ring Nebula. Furthermore, we would also need to understand why such 
clumps are only seen in the larger, more evolved PNe. Perhaps this 
is a selection effect related to the surface brightness. Spectroscopy 
and imaging with high spatial resolution of other PNe should 
provide more clues to the nature of the relation between recombination 
line enhancement and nebular diameter.

Unfortunately, we are not able to convincingly reconcile any of the
proposed models to explain the RL enhancements with all of the 
observations of NGC 6720 and NGC 6153. We present our results here 
in the hope that it will inspire new thinking on the physics behind 
emission processes in photoionized nebulae.

What are the consequences for abundance measurements from forbidden 
lines? The results are a bit mixed at the present time. The agreement
in abundances derived from optical/UV forbidden lines and IR lines
measured by $ISO$ for NGC 6153 \citep{liu00} offers strong evidence 
that temperature fluctuations do not have sufficient magnitude to
significantly affect forbidden-line abundances. On the other hand, 
there is a significant discrepancy between the Balmer discontinuity 
temperature and the [\oiii ] temperature in NGC 6153. It is possible 
that this discrepancy merely reflects variations in temperature across 
the nebula. As we have seen, the [\oiii ] temperature is not constant 
across the Ring Nebula, and such a variation is expected based on 
ionization models. A detailed observational study of the spatial 
profile of various temperature indicators across nebulae would 
provide a clearer picture of the temperature variation within
planetary nebulae. 

\acknowledgements

We thank A. Frank, J. Mathis, and G. Shields for very helpful and
stimulating discussions, Bob Benjamin for assistance in deriving 
the results shown in Figure 10, and the anonymous referee for 
several useful comments. The work of DRG is supported by NASA 
LTSA grant NAG5-7734, while HLD acknowledges support from NSF
grant AST 97-31156.

\clearpage

\begin{figure}
%\plotone{f1.ps}
\vspace{16.0cm}
\includegraphics{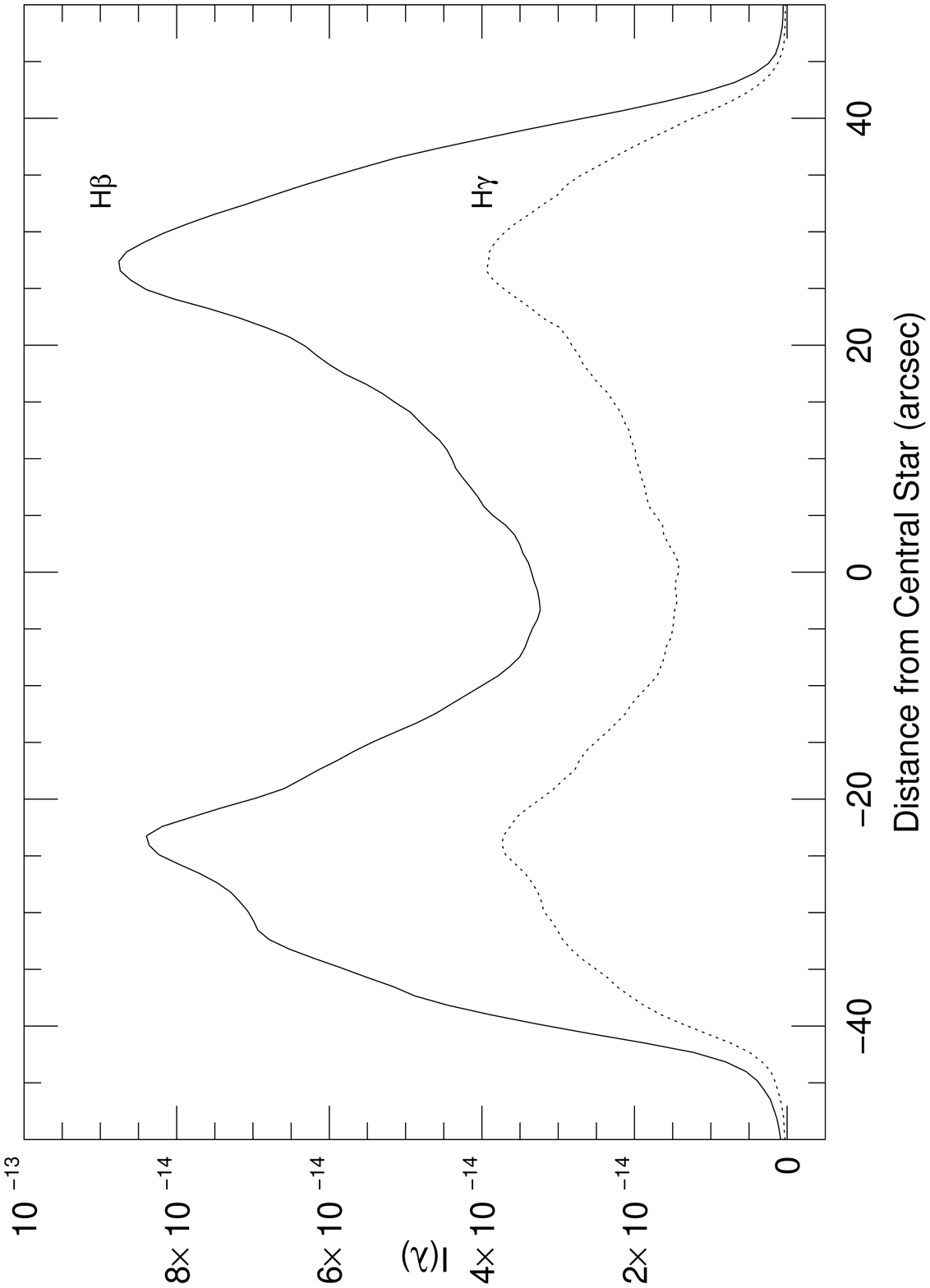}
\figcaption{Continuum-subtracted intensity profile of the H$\beta$ 
and H$\gamma$ emission lines along the slit in our spectrum of NGC 6720. 
East is to the left in this and subsequent similar figures. }
\end{figure}

\clearpage

\begin{figure}
%\plotone{f2.ps}
\vspace{16.0cm}
\includegraphics{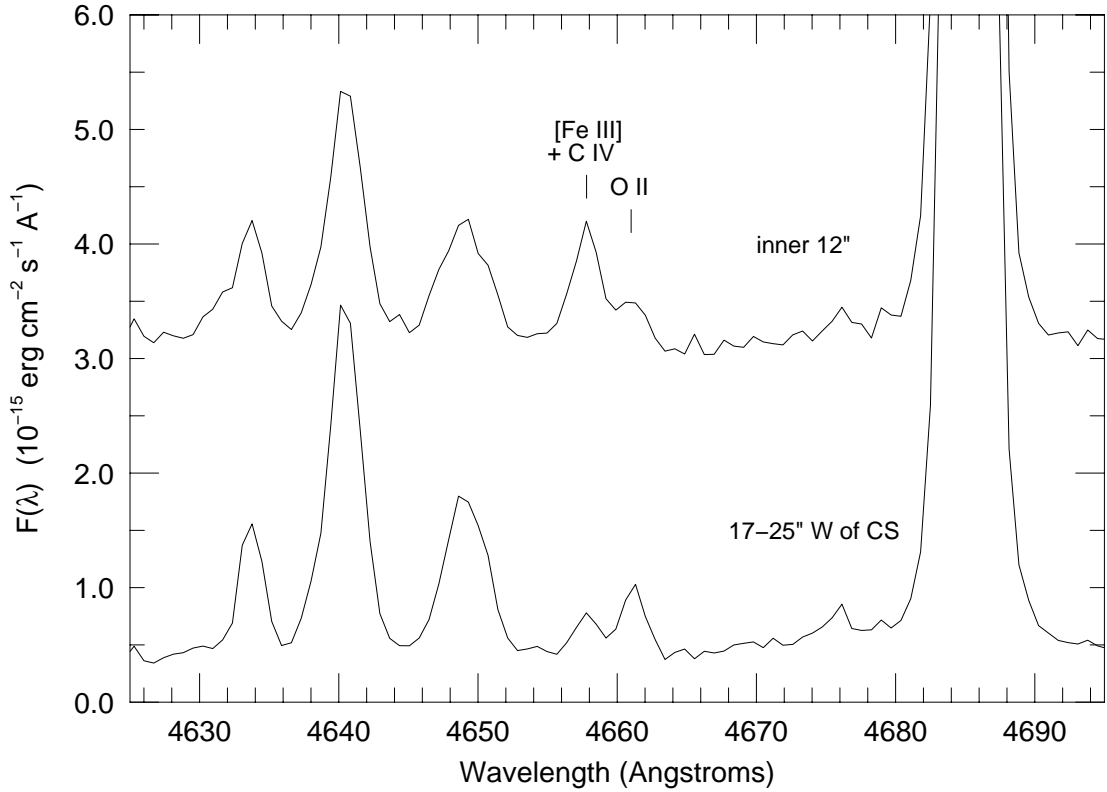}
\figcaption{Portions of extracted spectra around the \oii\ 4661 \AA\ line.
The upper spectrum is from the inner 12$\arcsec$ of the Ring Nebula, including
the central star. The lower spectrum is from a region 17-25$\arcsec$ west
of the central star, in the bright shell. Note that the ratio of [\feiii ]
$\lambda$4658 to \oii\ $\lambda$4661 reverses in the two spectra, but that
the two lines can be clearly distinguished in both cases.
}
\end{figure}
\clearpage 

\begin{figure}
%\plotone{f3.ps}
\vspace{16.0cm}
\includegraphics{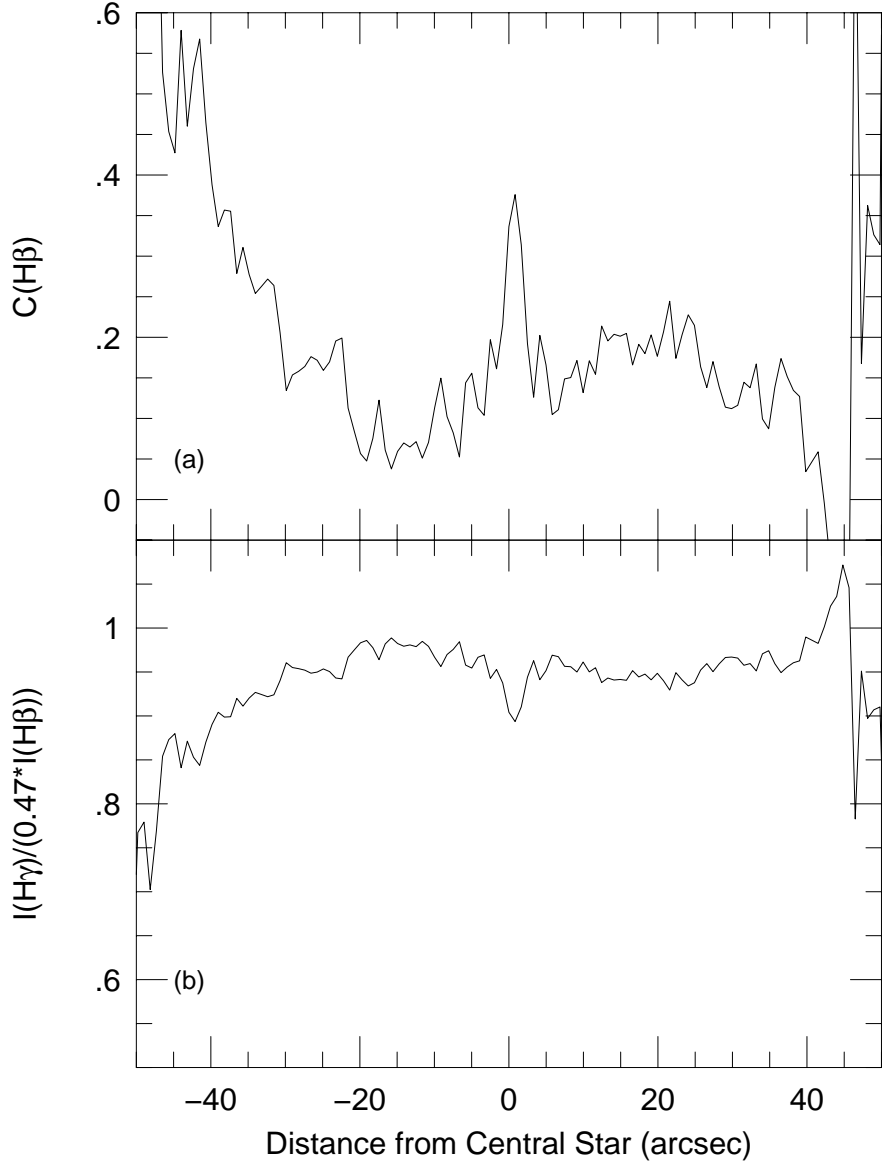}
\figcaption{(a) The spatial variation in reddening along our slit. C(H$\beta$)
is the logarithm extinction at H$\beta$. The spike at the center marks the
position of the central star, which likely has Balmer absorption lines 
that affect the emission line ratio. The steep upturn at the east end
and the steep downturn at the west end are likely due to small mismatches
in the alignment and focus for the H$\beta$ and H$\gamma$ profiles.
(b) The ratio I(H$\gamma$)/0.47$\times$I(H$\beta$) across NGC 6720. 
This ratio has a value of unity for $T_e$ $\approx$ 11,000 K; deviation 
of this ratio from unity is a measure of the interstellar reddening correction. 
This plot demonstrates that the reddening correction relative to 
H$\beta$ is less than 10\% along most of the slit for lines in our 
NGC 6720 spectrum.
}
\end{figure}

\clearpage 

\begin{figure}
%\plotone{f4.ps}
\vspace{16.0cm}
\includegraphics{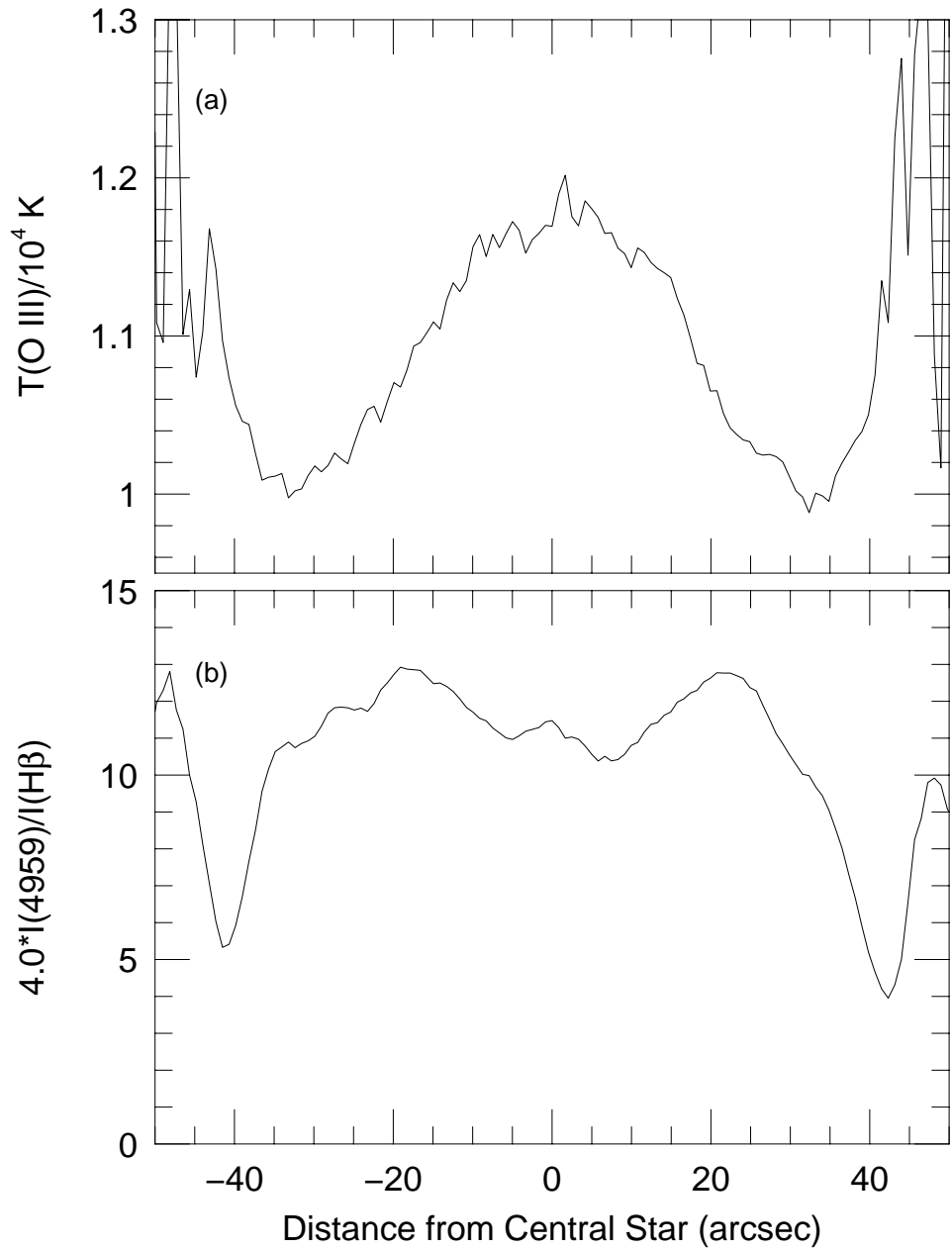}
\caption{(a) The variation of electron temperature across NGC 6720, 
derived from the [\oiii ] $\lambda$4959/$\lambda$4363 line ratio. 
Note that the T values for positions greater than $\pm$40$\arcsec$ from
the central star are unreliable because of poor signal/noise.
(b) The [\oiii ] (4959+5007)/H$\beta$ line ratio across NGC 6720.
}
\end{figure}

\clearpage 

\begin{figure}
%\plotone{f5.ps}
\vspace{16.0cm}
\includegraphics{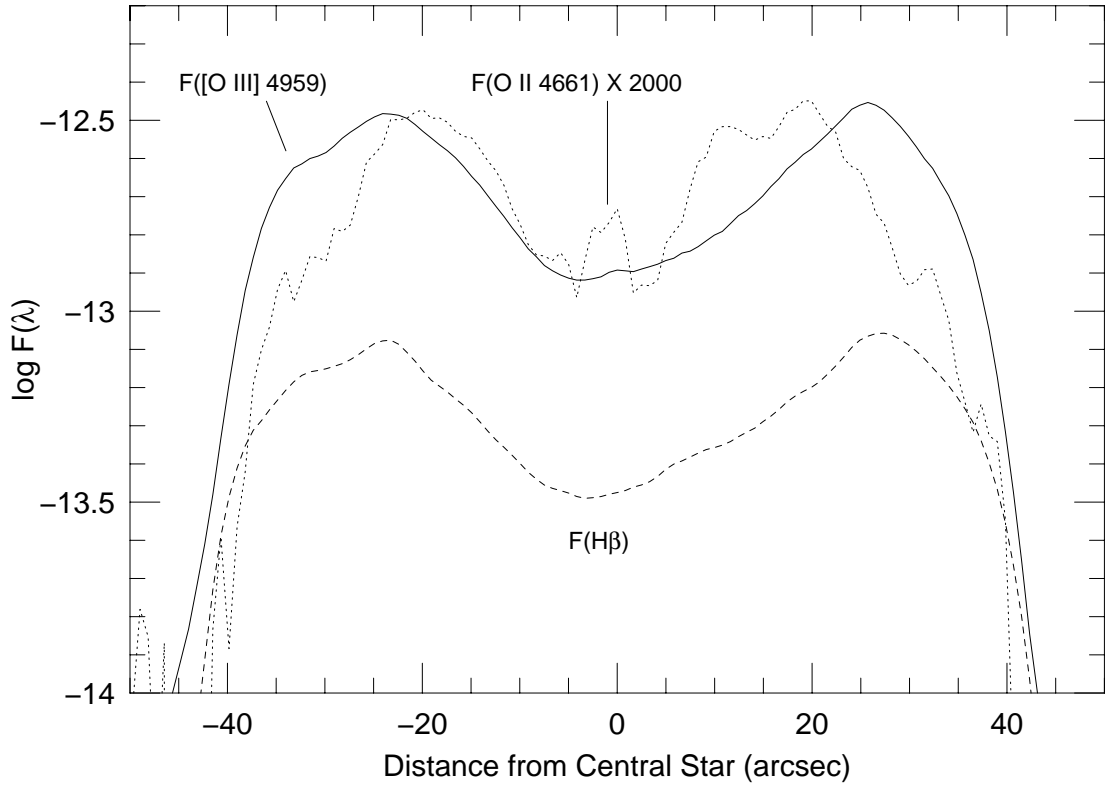}
\caption{The [\oiii ] (4959+5007) intensity profile along the slit (solid line) 
compared with \oii\ $\lambda$4661 (dotted line) and H$\beta$ (dashed line) in 
NGC 6720. Note that the \oii\ emission peaks interior to [\oiii ] and H$\beta$.
}
\end{figure}

\clearpage 

\begin{figure}
%\plotone{f6.ps}
\vspace{16.0cm}
\includegraphics{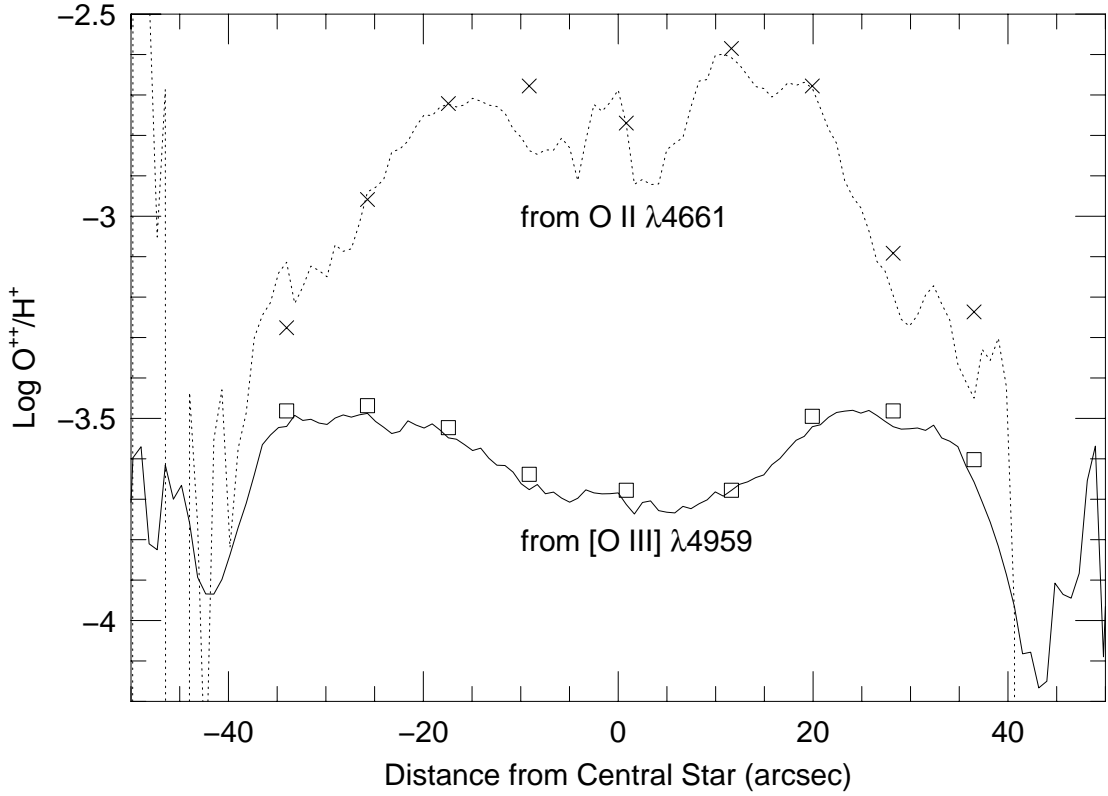}
\caption{The O$^{+2}$ abundance derived from [\oiii ] (solid line: from spatial
profile; squares: from binned extractions) and from \oii\ $\lambda$4661 
(dotted line: from spatial profile; crosses: from binned extractions). 
}
\end{figure}

\clearpage 

\begin{figure}
%\plotone{f7.ps}
\vspace{16.0cm}
\includegraphics{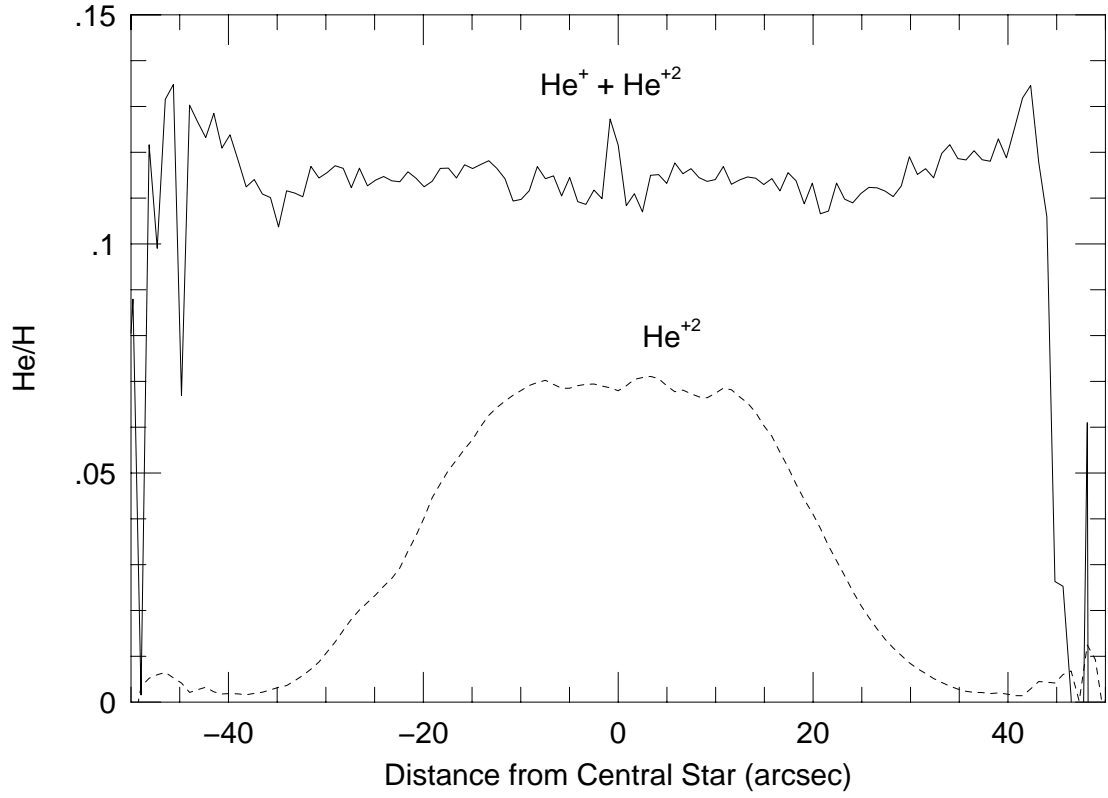}
\caption{The helium abundance relative to hydrogen across NGC 6720. The 
dashed line shows the contribution from He$^{+2}$. 
}
\end{figure}

\clearpage 

\begin{figure}
%\plotone{f8.ps}
\vspace{16.0cm}
\includegraphics{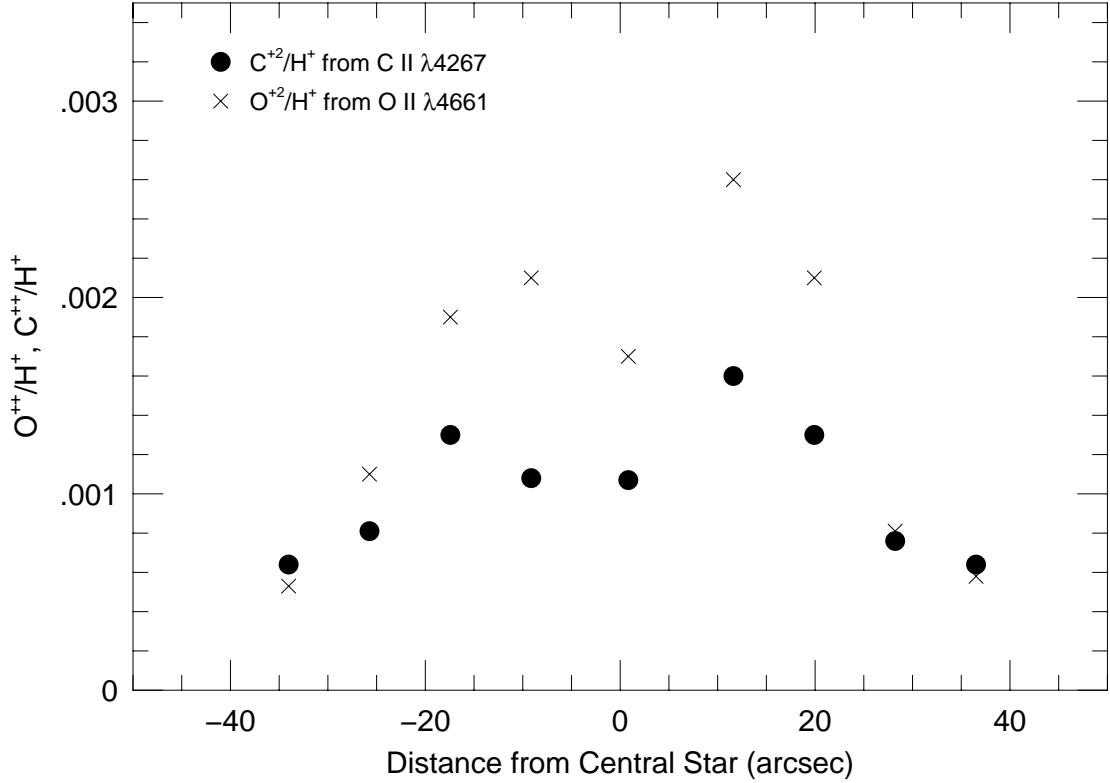}
\caption{The C$^{+2}$ abundance relative to hydrogen across NGC 6720, 
derived from the \cii\ 4267 \AA\ recombination line ({\it filled circles}). 
O$^{+2}$ abundances from \oii\ $\lambda$4661 are shown for comparison ({\it 
crosses}). The abundances were derived from the 
binned extractions as in Figure 5.
}
\end{figure}

\clearpage

\begin{figure}
%\plottwo{f9a.ps}{f9b.ps}
\vspace{18.5cm}
\includegraphics{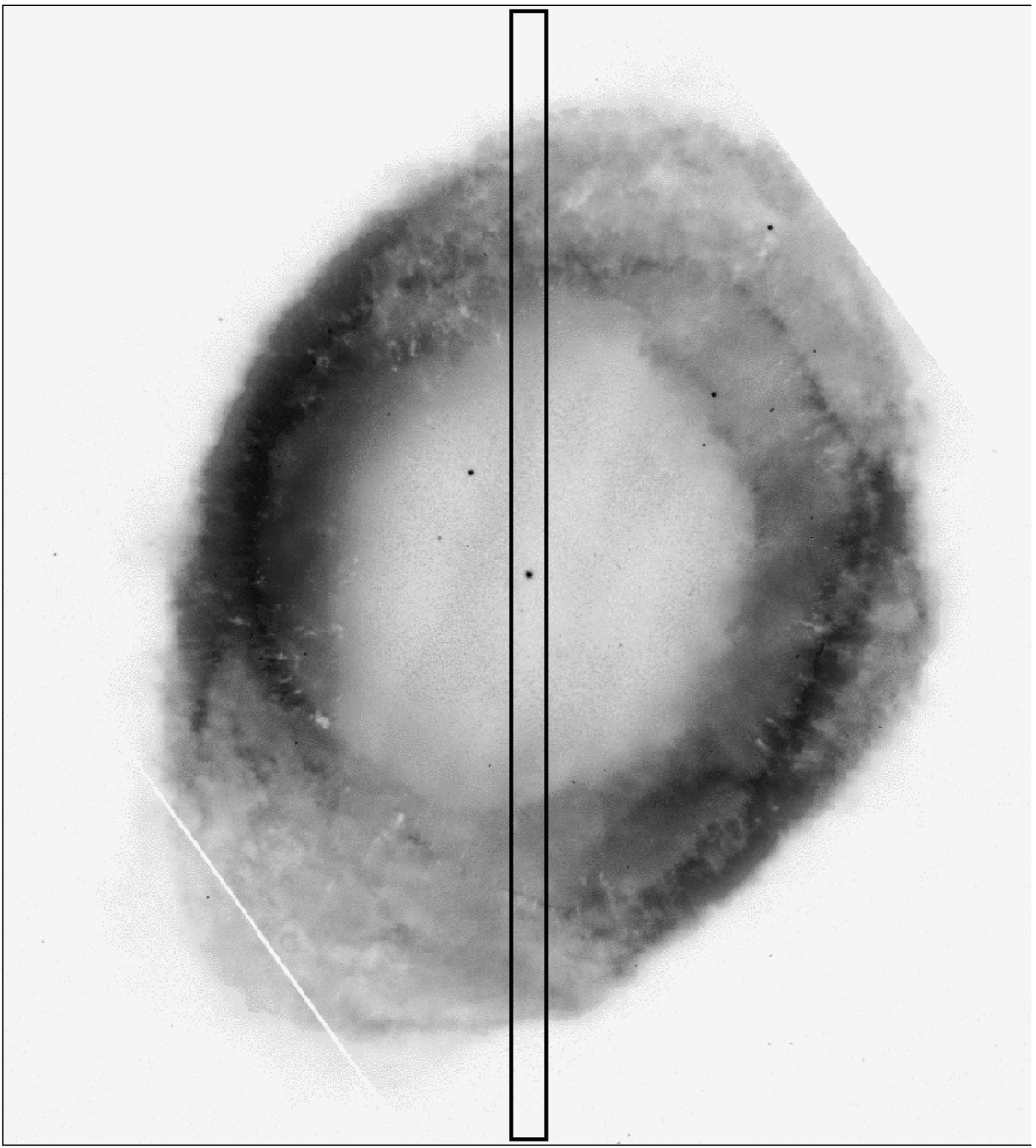}
\includegraphics{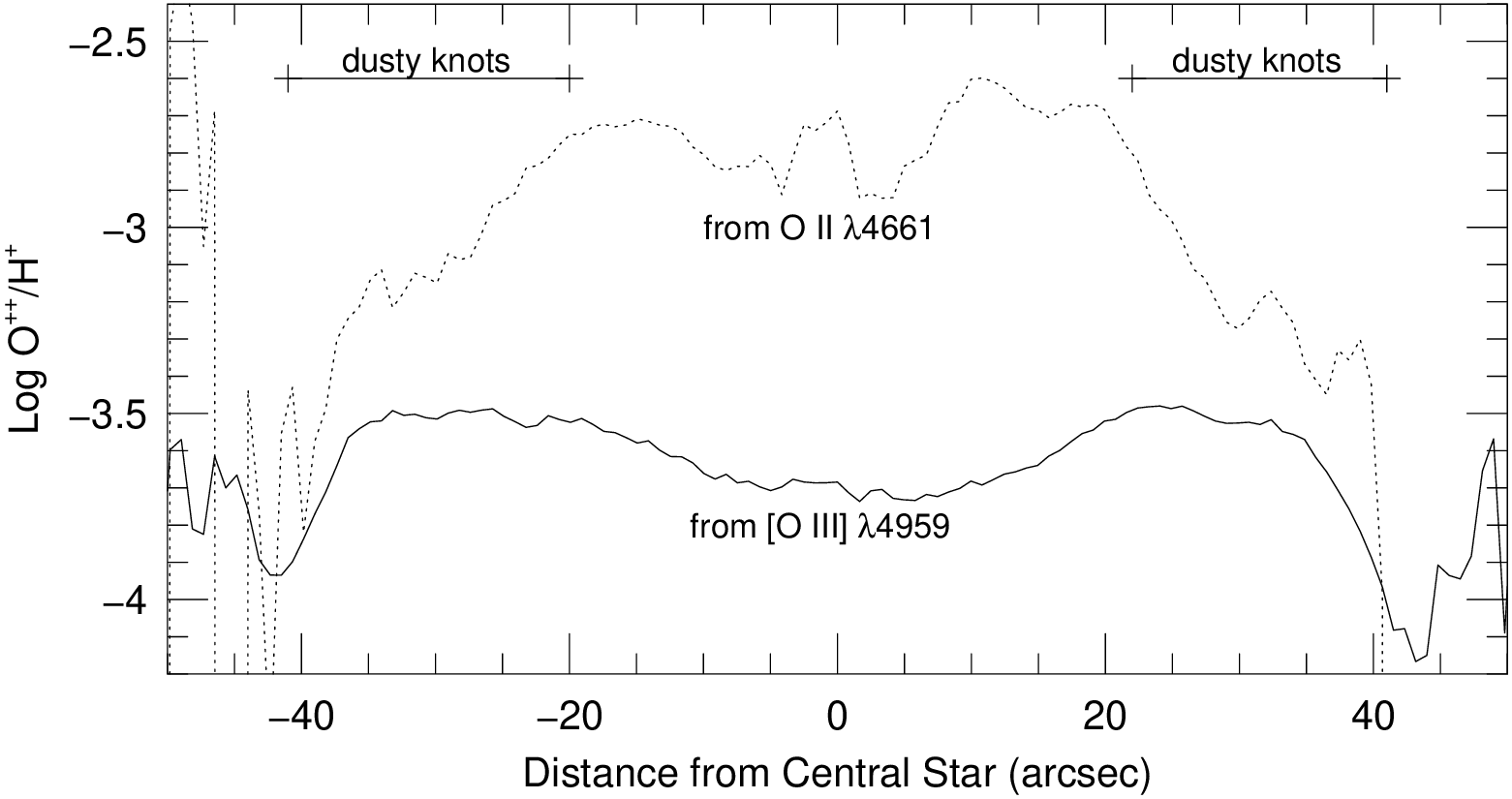}
\figcaption{
{\it Top:} A negative image of NGC 6720, showing numerous small dusty knots. 
The region of dust knots begins about 20$\arcsec$ from the central star (see 
scale in lower panel). The position and orientation of our spectrograph slit 
is marked. The image is a composite of WFPC-II images in the filters F469N,
F502N, and F658N.
{\it Bottom:} The O$^{+2}$ abundances derived from \oii\ and [\oiii ] 
across the Ring Nebula, as in Figure 5. The bars label the positions where 
dust knots are seen in the WFPC-2 images. The top and bottom panels 
have the same horizontal spatial scale. 
}
\end{figure}

\clearpage 

\begin{figure}
%\plotone{f10.ps}
\vspace{16.0cm}
\includegraphics{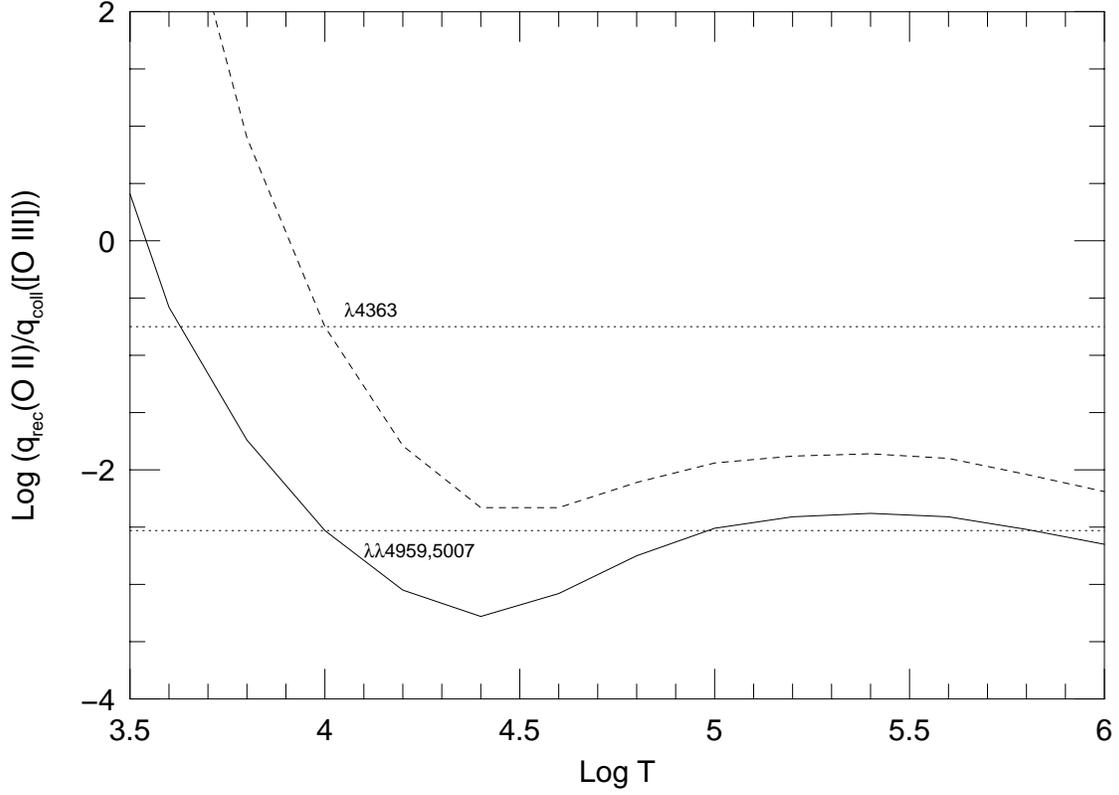}
\caption{The ratio of the recombination rate $q_{rec}$ to the collisional
excitation rate $q_{coll}$ for [\oiii]. The solid curve shows the ratio 
with respect to [\oiii] $\lambda\lambda$4959,5007, while the dashed curve
shows the ratio with respect to [\oiii] $\lambda$4363. The dotted lines
mark the ratios evaluated at 10,000 K, the electron temperature derived
for the shell of NGC 6720.
}
\end{figure}

\end{document}